\documentstyle[12pt]{article}

\begin{document}

\font\mybb=msbm10 at 12pt
\def\bb#1{\hbox{\mybb#1}}
\def\Z {\bb{Z}}
\def\R {\bb{R}}

\begin{titlepage}

\begin{flushright}
UG-7/95\\
QMW-PH-95-23\\
{\bf hep-th/9508091}\\
August $18$th 1995
\end{flushright}

\begin{center}

{\Large {\bf $S$ Duality and Dyonic $p$--Brane Solutions in Type~II
String Theory}} \vspace{.5cm}

{\bf Eric Bergshoeff\footnote{E-mail: {\tt bergshoe@th.rug.nl}} and
Harm Jan Boonstra\footnote{E-mail: {\tt boonstra@th.rug.nl}}}

{\it Institute for Theoretical Physics, University of Groningen}\\
{\it Nijenborgh 4, 9747 AG Groningen, The Netherlands}
\vspace{.5cm}

{\bf Tom\'as Ort\'{\i}n}\footnote{E-mail: {\tt
t.ortin@qmw.ac.uk}\\Address after October 1st: CERN Theory
Division, CH-1211, Gen\`eve 23, Switzerland.}

{\it Department of Physics, Queen Mary \& Westfield College}\\
{\it Mile End Road, London E1 4NS, U.K.}
\end{center}


\vspace{.3cm}

\begin{abstract}

We show how a solitonic ``magnetically'' charged $p$-brane solution of a
given supergravity theory, with the magnetic charge carried by an
antisymmetric tensor gauge field, can be generalized to a dyonic
solution.  We discuss the cases of ten-dimensional and
eleven-dimensional supergravity in more detail and a new dyonic
five-brane solution in ten dimensions is given.  Unlike the purely
electrically or magnetically charged five-brane solution the dyonic
five-brane contains non-zero Ramond--Ramond fields and is therefore an
intrinsically type~II solution.  The solution preserves half of the
type~II spacetime supersymmetries.  It is obtained by applying a
solution-generating $SL(2,\R) \times SL(2,\R)$ $S$~duality
transformation to the purely magnetically charged five-brane solution.
One of the $SL(2,\R)$ duality transformations is basically an extension
to the type~II case of the six-dimensional $\Z_2$ string/string duality.

We also present an action underlying the type IIB supergravity theory.

\end{abstract}

\end{titlepage}

\newpage

\pagestyle{plain}


\section*{Introduction}

Recently, an active field of research has been the search for $p$-brane
solutions ({\it i.e.}~solutions with $p$ translational spacelike
isometries) in supergravity theories (for some recent reviews, see {\it
e.g.}~\cite{Ca1,Ho1,Du1}).  One of the motivations is that, in case the
supergravity theory is an effective superstring theory, they might give
us information about the strong-coupling behaviour of the superstring.
Two particularly interesting examples corresponding to the
ten-dimensional heterotic string effective action are the fundamental
string solution found by Dabholkar {\it et al.}~\cite{Da1} and the
five-brane soliton found by Strominger \cite{St1}.  Many more $p$-brane
solutions have been found, both in $D=10$ \cite{Ca1,Ho1,Du1,Ho2,Du2} as well as
in $D=11$ \cite{Du3}.

Most of the $p$-brane solutions found so far are either elementary
solutions requiring a singular source term (that we will call
``electrically charged'') or solitonic solutions (``magnetically
charged'').  It is natural to investigate the possibility of
constructing {\it dyonic} $p$-brane solutions.  Sofar, most of the
research has been concentrating on the study of
four-dimensional dyonic black holes ($0$-branes), see
{\it e.g.}~\cite{Sh1,Cv1}.  In this work we will consider
this problem for higher $p$ and higher dimensions\footnote{
We will construct solutions with nonzero Ramond-Ramond fields. Since
the Ramond-Ramond fields do not couple via a standard sigma model action
to the type II string it is nontrivial to deal with source terms. In this
paper we postpone a proper treatment of this issue and only solve the
source-free equations of motion.}.  To explain the basic
idea we start by succinctly describing the ten-dimensional five-brane
electric-magnetic (e-m) $\Z_2$ map.  Any ten-dimensional heterotic
five-brane solution $5_{(10)}$ can be reinterpreted via dimensional
reduction as a six-dimensional string ($1$-brane) solution $1_{(6)}$.
If the original $5_{(10)}$ solution does not give rise to
six-dimensional vector fields, then one can use the string/string
duality symmetry \cite{Du4,Du5} to generate another six-dimensional
solution $1_{(6)}^{\prime}$ which in turn can be reinterpreted via
``inverse dimensional reduction'' as another ten-dimensional five-brane
solution $5_{(10)}^{\prime}$.  Since the six-dimensional transformation
is an e-m duality transformation on the axion two-form, we get a {\it
ten-dimensional e-m duality for heterotic five-branes}.

As an example, let us consider, the magnetic $5_{(10){\rm m}}$
solution\footnote{Our conventions are those of Ref.~\cite{Be2}.  We only
discuss the so-called ``neutral five-brane'' \cite{Du6,Ca2}.  This
example has previously been studied in \cite{Du8,Se1,Ha1}.} \cite{St1} in
the string frame\footnote{The Einstein frame metric ${\hat
g}_{\hat\mu\hat\nu}^{E}$ and the string frame metric ${\hat
g}_{\hat{\mu}\hat{\nu}}^{S}$ in ten dimensions are related by ${\hat
g}_{\hat \mu\hat \nu}^{E} = e^{-{\textstyle\frac{1}{2}}\hat\phi}{\hat
g}_{\hat\mu\hat\nu}^{S}$.}:

\begin{equation}
{\bf 5_{(10){\rm m}}}\left\{ \begin{array}{rcl} ds^2 &=&
(dx^0)^2 - (dx^1)^2 - (dx^m)^2 - e^{2\hat{\phi}}dx^adx^a\, ,\\
\hat{H}_{abc} &=& {\textstyle\frac{2}{3}}\epsilon_{abcd}\partial^d
\hat{\phi}\, , \label{5brane}
\end{array}\right.
\end{equation}

\noindent where $x^{\hat{\mu}}=(x^0,x^1,x^m;x^a), m\in \{6,7,8,9\},
a\in\{2,3,4,5\}$.  The five-brane world-volume is parametrized by
$\{x^0,x^1,x^m\}$.  The dilaton only depends on the $x^a$'s and
satisfies $\Box e^{2\hat{\phi}}=0$ where
$\Box=\delta^{ab} \partial_a\partial_b$.  This solution can be
reinterpreted as a string in six dimensions by eliminating the
$x^{m}$'s:

\begin{equation}
{\bf 1_{(6){\rm m}}}
\left\{
\begin{array}{rcl}
ds^2 &=& (dx^0)^2-(dx^1)^2
 - e^{2{\phi}}(dx^a)^2\, ,\\
{H}_{abc} &=& {\textstyle\frac{2}{3}}\epsilon_{abcd}\partial^d {\phi}\,
, \label{1m}
\end{array}\right.
\end{equation}

\noindent where the dilaton satisfies $\Box e^{2\phi}
=0$\footnote{We express all solutions in terms of their own dilatons.}.
Now we make use of the six-dimensional $\Z_{2}$ e-m duality.
This is most easily described in the Einstein frame\footnote{The
Einstein frame metric $g_{\mu\nu}^{E}$ and the string frame metric
$g_{\mu\nu}^{S}$ in six dimensions are related by $g_{\mu\nu}^{E} =
e^{-\phi}g_{\mu\nu}^{S}$.} by

\begin{equation}
\phi^{\prime} =  -\phi\,,\hskip 1truecm
H^{\prime}  =   e^{-2\phi}{}^{\  *}H\,,
\label{Z2}
\end{equation}

\noindent with $*$ the six-dimensional Hodge star.  In the string frame
this transformation acts on the metric.  The result is the {\it
six-dimensional fundamental string solution}

\begin{equation}
{\bf 1_{(6){\rm e}}}
\left\{
\begin{array}{rcl}
ds^2 &=& e^{2{\phi}}\left [(dx^0)^2-(dx^1)^2\right ]
 -(dx^a)^2\, ,\\
       {B}_{01}&=& e^{2{\phi}}\, .
\label{1e}
\end{array}\right.
\end{equation}

\noindent Observe that the dilaton now satisfies $\Box e^{-2\phi}=0$.
Finally, we can reinterpret this solution as a five-brane solution
in ten dimensions:\footnote{
The $5_{(10)\rm e}$ solution can be viewed as a special case of the
Dabholkar string $1_{(10)\rm e}$ where the eight-dimensional Laplace
equation for the dilaton has been solved in the presence of four extra
isometries.  As a consequence, the Dabholkar string has an
eight-dimensional spherical symmetry while the $5_{(10)\rm e}$ solution
has a four-dimensional spherical symmetry.}

\begin{equation}
{\bf 5_{(10)\rm e}}
\left\{
\begin{array}{rcl}
ds^2 &=& e^{2\hat{\phi}}\left [(dx^0)^2-(dx^1)^2\right ]
 -(dx^m)^2 -(dx^a)^2\, ,\\
       \hat{B}_{01}&=& e^{2\hat{\phi}}\,.
\label{FS}
\end{array}\right.
\end{equation}

\noindent The dilaton depends only on the $x^a$'s and satisfies
$\Box e^{-2\hat\phi}=0$.  This solution can be thought
of as an electric ten-dimensional heterotic five-brane solution which
is the e-m dual of the solitonic five-brane $5_{(10){\rm m}}$.

It is natural to try to generalize the e-m duality $\Z_2$ to $SL(2,\R)$
so that we can use it to build dyonic solutions.  In doing so we
encounter the following two problems.

\begin{enumerate}

\item We have restricted ourselves to the case in which no
six-dimensional vector fields arise.  Otherwise, in the framework of the
heterotic string compactified on $T^{4}$, the dual theory does not
coincide with the original theory.  A way of seeing this is the
following.  In the presence of vector fields there are Chern-Simons
terms in the field-strength for the axion.  In dualizing the theory,
these Chern-Simons terms get interchanged with topological terms in the
action and, vice versa, every topological term in the action gives rise
to a Chern-Simons term\footnote{For instance, in four dimensions, the
axion field strength has Chern-Simons terms so $H=\partial
B+\frac{1}{2}AF$ and there are no topological terms present in the
action.  In the dual theory the axion is substituted by the pseudoscalar
$a$ and the topological term $aF{}^{\star}F$ appears in the action.}.
Therefore, the duality just described can only be a symmetry of the
equations of motion if for every Chern-Simons term there is a
corresponding topological term.  However, in the heterotic case there
are no topological terms whatsoever.  Therefore, the duality just
described would not be a symmetry of the equations of motion of a single
theory, but would relate two different (dual string) theories.

\item To extend the $\Z_{2}$ symmetry to $SL(2,\R)$ one needs more
fields coupled to the axion in a specific way.  This is similar to the
four-dimensional case where, in the absence of the pseudoscalar axion
$a$, one only has a $\Z_{2}$ e-m duality.  The introduction of $a$
suitably coupled to the vectors enhances the symmetry to $SL(2,\R)$.

\end{enumerate}

The solution to both of these problems lies in the type~II theories: In
the type~IIB theory there are topological terms present from the
beginning, and they are such that they get interchanged with the
Chern-Simons terms that appear in the compactification, leaving the
theory invariant.  At the same time, in these topological terms the axion
is coupled to the four-form $\hat{D}$ (which is a RR field) in the
``right way'', so we can extend the above five-brane $\Z_2$ to an
$SL(2,\R)$ and we can build dyonic five-branes.  The price to
pay is that, although we can always start with a type~I five-brane, in
general we will obtain type~II five-branes with extra non-zero RR fields
which seem to be necessary for the symmetry enhancement (see, however,
the conclusions).

In the following section we will briefly review type~IIB supergravity.
As a new result, which is useful for our present purposes, we will
present an action underlying the type~IIB theory.  In the next section
we will dimensionally reduce this action using an ansatz simple but rich
enough to show the symmetry enhancement mechanism.  In section 3 we will
obtain a new ten-dimensional type~II dyonic five-brane solution which
continually interpolates between the $5_{(10){\rm m}}$ and $5_{(10){\rm
e}}$ solutions given in (\ref{5brane}) and (\ref{FS}), respectively.
Finally, in the conclusions we will discuss more general applications of
our techniques.

Part of the results of this letter have been presented in
\cite{Be1}.


\section{$D=10$ Type~IIB Supergravity}
\label{sec-iibaction}

It is known \cite{Mar1} that the field equations of $D=10$ type~IIB
supergravity \cite{Sc1} cannot be derived from a covariant action.
Nevertheless, it is useful to think about an ``action'' in the
restricted sense explained below.  The only equation of motion that
cannot be obtained from an action is that of the four-form gauge field
$\hat{D}$.  This equation of motion states that the field strength
$\hat{F}$ of $\hat{D}$ is self-dual: $\hat{F}={}^{\star}\hat{F}$.  It
follows that if one sets $\hat{F}=0$ everywhere in the equations of
motion, one should be able to obtain the resulting reduced set of
equations from an action, by varying with respect to all fields but
$\hat{D}$.  This was done in Ref.~\cite{Be2}.

We will show in this section that one can even write down an action
involving ${\hat{F}}$.  A useful property of such an action is that,
when properly used, it leads to the correct action for the {\it
dimensionally reduced} type~IIB supergravity theory.  We thus may avoid
the dimensional reduction of the ten-dimensional type~IIB field
equations which is more complicated.  This property will be exploited in
the next section.

The idea is the following: we keep $\hat{F}$ different from zero but
eliminate the self-duality constraint.  Of course, we would like to have
an equation of motion for $\hat{D}$ replacing
$\hat{F}={}^{\star}\hat{F}$.  As a matter of fact, there is a perfect
``spare equation of motion'' at our disposal.  One of the consequences
of the self-duality constraint is that the equation of motion of
$\hat{D}$ is equal to its Bianchi identity.  Therefore, it is natural to
take as equation of motion for $\hat{D}$\footnote{The term at the r.h.s.
follows from the Chern--Simons term in $\hat F$, see below.}

\begin{equation}
\nabla_{\hat{\mu}}\hat{F}^{\hat{\mu}\hat{\mu}_{1}
\hat{\mu}_{2} \hat{\mu}_{3} \hat{\mu}_{4} }=
{\textstyle\frac{3}{5!4}}\epsilon^{ij}
\epsilon^{\hat{\mu}_{1}\ldots\hat{\mu}_{10}}
\hat{\cal H}^{(i)}_{\hat{\mu}_{5}\hat{\mu}_{6}\hat{\mu}_{7}}
\hat{\cal H}^{(j)}_{\hat{\mu}_{8}\hat{\mu}_{9}\hat{\mu}_{10}}\, .
\label{eq:Dneweq}
\end{equation}

\noindent This equation is compatible with the self-duality constraint
(which we have temporarily abolished), but it does not imply it.  So, in
fact, we just have eliminated the self-duality constraint in a
consistent way.

Now, is there an action for the new set of equations of motion obtained
by eliminating the self-duality constraint and substituting it by
Eq.~(\ref{eq:Dneweq})?  Note that in the original equations, the
self-duality of $\hat{F}$ was already taken into account and therefore
only $\hat F$ occurs.  In this non-self-dual (NSD) theory we expect both
$\hat{F}$ and ${}^{\star}\hat{F}$ to occur.  The NSD theory is defined
by the property that it has the same field content as the original
theory but $\hat{F}$ is not self-dual and, if one imposes self-duality
in the field equations, one recovers it.

It turns out that the easiest way to find the NSD theory and its action
is to make the most obvious ansatz for its action: add to the action for
the $\hat{F}=0$ case an $\hat{F}^{2}$ (kinetic) term and a topological
term with numerical factors to be adjusted.  One easily finds that the
action we are looking for, in the string frame and with the notation and
conventions of Ref.~\cite{Be2}, is given by\footnote{
In this paper we will use the convention that $\hat\varphi$ is the
dilaton and $\hat {\cal B}^{(1)}$ the NS-NS axion. Note that other definitions
of the dilaton and NS-NS axion are possible which differ from
$\hat\varphi$ and $\hat {\cal B}^{(1)}$ by an $SL(2,\R)$ rotation.}:

\begin{eqnarray}
\label{eq:stringynonselfdualaction}
\hat{S}_{\rm NSD-IIB}^{{\rm string}} & = &
{\textstyle\frac{1}{2}} \int d^{10}x
\sqrt{-\hat{\jmath}}
\left\{ e^{-2\hat{\varphi}}
\left[ -\hat{R} \left( \hat{\jmath} \right)
+4(\partial\hat{\varphi})^{2} -{\textstyle\frac{3}{4}}
\left(\hat{\cal H}^{(1)} \right)^{2} \right] \right.
\nonumber \\
& &
\\
& & \left.  -{\textstyle\frac{1}{2}} (\partial\hat{\ell})^{2}
 -{\textstyle\frac{3}{4}} \left( \hat{\cal H}^{(2)} -\hat{\ell}
\hat{\cal H}^{(1)} \right)^{2}
-{\textstyle\frac{5}{6}\hat{F}^{2}}-
{\textstyle\frac{1}{96{\sqrt {-\hat\jmath}}}}\epsilon^{ij}\epsilon\hat{D}
\hat{\cal H}^{(i)}\hat{\cal H}^{(j)}
\right\}\, .\nonumber
\end{eqnarray}

\noindent For the sake of completeness we list the definitions of
the field strengths and gauge transformations for the type~IIB fields
$\bigl\{\hat{D}_{\hat{\mu}\hat{\nu}\hat{\rho}\hat{\sigma}},\
\hat{\jmath}_{\hat{\mu}\hat{\nu}},\ \hat{\cal
B}^{(i)}_{\hat{\mu}\hat{\nu}},\ \hat{\ell},\ \hat{\varphi}\bigr\}$,
$(i=1,2)$:

\begin{equation}
\begin{array}{rclrcl}
\hat{\cal H}^{(i)}
&
=
&
\partial\hat{\cal B}^{(i)}\, ,
&
\delta\hat{\cal B}^{(i)}
&
=
&
\partial\hat{\Sigma}^{(i)}\, ,
\\
& & & & &
\\
\hat{F}
&
=
&
\partial\hat{D} +{\textstyle\frac{3}{4}}
\epsilon^{ij}\hat{\cal B}^{(i)} \partial\hat{\cal B}^{(j)}\, ,
&
\delta\hat{D}
&
=
&
\partial\hat{\rho} -{\textstyle\frac{3}{4}}
\epsilon^{ij}\partial\hat{\Sigma}^{(i)} \hat{\cal B}^{(j)}\, .
\end{array}
\end{equation}

Varying with respect to all the fields one gets the equations motion of
the NSD theory and, imposing the self-duality constraint, these become
the equations of the type~IIB theory.

The NSD theory defined by Eq.~(\ref{eq:stringynonselfdualaction}) has
all the symmetries of the type~IIB theory, including the global
$SL(2,\R)_{\rm IIB}$, and an additional global $\Z_2$ e-m duality of the ${\hat
D}$
field that interchanges $\hat{F}$ and ${}^{\star}\hat{F}$.  To exhibit
the $SL(2,\R)_{\rm IIB}$ symmetries, it is useful to go to the Einstein
frame because the Einstein metric is inert under them:

\begin{eqnarray}
\label{IIB-E}
S_{\rm NSD-IIB}^{\rm Einstein}
& = &
{\textstyle\frac{1}{2}}\int d^{10}x\ \sqrt{-\hat{g}}\
\left\{
-\hat{R} +{\textstyle\frac{1}{4}}
{\rm Tr}\left(\partial_{\mu}\hat{\cal M}
\partial^{\mu}\hat{\cal M}^{-1}\right)
-{\textstyle\frac{3}{4}}
\tilde{\hat{\cal H}}_{(i)}\hat{\cal H}^{(i)}
\right.
\nonumber \\
& &
\nonumber \\
& &
\left.
-{\textstyle\frac{5}{6}\hat{F}^{2}}-
{\textstyle\frac{1}{96{\sqrt {-\hat{g}}}}}\epsilon^{ij}\epsilon\hat{D}
\hat{\cal H}^{(i)}\hat{\cal H}^{(j)}
\right\}\, ,
\label{IIBactionE}
\end{eqnarray}

\noindent where $\hat{g}_{\hat{\mu}\hat{\nu}}
=e^{-\frac{1}{2}\hat{\varphi}} \hat{\jmath}_{\hat{\mu}\hat{\nu}}$ is the
Einstein-frame metric, $\hat{\cal M}$ is the $2\times 2$ matrix

\begin{equation}
\hat{\cal M}=
\left(
\hat{\cal M}_{ij}
\right)
=
\frac{1}{\Im {\rm m}\lambda}
\left(
\begin{array}{rr}
|\hat{\lambda}|^{2}        & -\Re {\rm e}\hat{\lambda} \\
-\Re {\rm e}\hat{\lambda}  &               1           \\
\end{array}
\right)\, ,
\end{equation}

\noindent where $\hat{\lambda}=\hat{\ell}+ie^{-\hat{\varphi}}$ is a complex
scalar that parametrizes $SL(2,\R)_{\rm IIB}$ and

\begin{equation}
\tilde{\hat{\cal H}}_{(i)}=\hat{\cal H}^{(j)}\hat{\cal M}_{ji}\, .
\end{equation}

The action (\ref{IIB-E}) is invariant under the $SL(2,\R)_{\rm IIB}$
transformations

\begin{eqnarray}
\hat{\cal H}^{\prime} & = & \Lambda\hat{\cal H}\, ,
\nonumber \\
\hat{\cal M}^{\prime} & = & \left( \Lambda^{-1}\right)^{T}
\hat{\cal M}\Lambda^{-1}\, .
\label{eq:sl2r}
\end{eqnarray}

\noindent If $\Lambda$ is the $SL(2,\R)_{\rm IIB}$ matrix

\begin{equation}
\Lambda=
\left(
\begin{array}{rr}
 a & -c \\
-b &  d \\
\end{array}
\right)\, ,
\end{equation}

\noindent the transformation Eq.~(\ref{eq:sl2r}) of the matrix $\hat{\cal
M}$ implies the usual transformation of the complex scalar
$\hat{\lambda}$

\begin{equation}
\hat{\lambda}^{\prime}=\frac{a\hat{\lambda}+b}{c\hat{\lambda}+d}\, .
\end{equation}

Although this symmetry does not involve any e-m duality rotation
$\hat {\cal H} \rightarrow {}^*\hat {\cal H}$ in the ten-dimensional
space-time, from the string theory point of view it is a genuine
$S$~duality symmetry \cite{Hu2} since some of its transformations
($\hat{\lambda}^{\prime}=-1/\hat{\lambda}$) interchange the strong- and
weak-coupling regimes of string theory and world-sheet elementary
excitations (NS-NS states) with solitons (RR states) \cite{Be2}.
In fact, as we will see in section 3, it is the $SL(2,\R)_{\rm IIB}$
which transforms electric solutions into magnetic solutions and
which can be used by itself to construct dyonic solutions. However,
these dyonic solutions are restricted in the sense that if the
electric (magnetic) charge is carried by the NS-NS axion then the
magnetic (electric) charge is carried by the RR axion. In the next
section we will introduce another $SL(2,\R)$ transformation, which
we will call $SL(2,\R)_{\rm EM}$, and which will enable us to
construct dyonic solutions where both axions carry an electric as well
as a magnetic charge.

Finally, we expect that a NSD type~IIB theory including fermions can also be
found.  Of course, the full NSD type~IIB action cannot be
supersymmetric.  However, the supersymmetry should be recovered in the
field equations when the (super-) self-duality contraint is
imposed\footnote{We thank M.~Green for a discussion on this point.}.


\section{Type~II~EM duality in Six Dimensions}
\label{sec-compac}

In this section we will reduce the NSD type~IIB action
(\ref{eq:stringynonselfdualaction}) to six dimensions using the
following simplified ansatz for the fields ($\mu,\nu,...$ are
six-dimensional spacetime indices and $m,n,...$ are the four internal
directions)\footnote{It turns out that not $\bar \varphi,\bar G$ but
combinations of them for which we reserve the names $\varphi, G$,
naturally fit into a $SL(2,\R)$ coset.}:

\begin{equation}
\label{ansatzspecial}
\begin{array}{rclrcl}
\hat{\jmath}_{\mu\nu} & = &\jmath_{\mu\nu}\, , &
\hat{\jmath}_{mn} & = & -e^{\bar{G}}\delta_{mn}\,,\\
\hat{{\cal B}}^{(i)}_{\mu\nu} & = & {\cal B}^{(i)}_{\mu\nu}\, ,
& & & \\
\hat{D}_{\mu\nu\rho\sigma}& = & D_{\mu\nu\rho\sigma}\, ,&
\hat{D}_{mnpq}& = & D_{mnpq}\, .\\
\hat{\ell} & = & \ell\, , &
\hat{\varphi} & = & \bar{\varphi}+\bar{G}\, . \\
\end{array}
\end{equation}

\noindent All other components are zero.  This ansatz contains extra
scalars (as compared to the ansatz used in the example of the
introduction) which are the RR fields necessary to extend $\Z_2$
to $SL(2,\R)$ as we will show below.  Note that for the $D$
field we only consider the scalars that arise in six dimensions
consistently with self-duality.  This gives precisely one scalar:
$D=\epsilon^{mnpq}D_{mnpq}$.  In the dimensional reduction we dualize
the field strength $F(D)_{\mu_1 ...\mu_5}$ to the field strength of a
scalar $\tilde{D}$ and a suitable normalization turns the self-duality
constraint into $\tilde{D}=D$, which can now be substituted into the
action.  The kinetic and topological terms for $D$ and $\tilde{D}$ then
give equal contributions to the reduced action.  Therefore it suffices
to collect the $D_{mnpq}$ terms only and multiply these terms by a
factor two.

\noindent The resulting reduced action is

\begin{eqnarray}
S={\textstyle\frac{1}{2}}\int d^6 x\sqrt{-\jmath}\biggl [e^{-2\varphi}
  \bigl [-R(\jmath)
  +4(\partial\varphi)^2 - (\partial\bar{G})^2
-{\textstyle\frac{3}{4}}({\cal H}^{(1)})^2
  \bigr ]\nonumber\\
 -{\textstyle\frac{1}{2}}e^{2\bar{G}}(\partial\ell)^2
  -{\textstyle\frac{3}{4}}e^{2\bar{G}}({\cal H}^{(2)}-\ell{\cal
H}^{(1)})^2 \\
 -{\textstyle\frac{1}{72}}e^{-2\bar{G}}(\partial D)^2
  +{\textstyle\frac{1}{8}}D{\cal H}^T{\cal L}{}^{\ *}{\cal H}\biggr
]\nonumber\, ,
\end{eqnarray}

\noindent where $({}^{*}{\cal H})_{\mu\nu\rho} =
{\textstyle\frac{1}{6\sqrt{-j}}}
\epsilon_{\mu\nu\rho\alpha\beta\gamma}{\cal
H}^{\alpha\beta\gamma}$ and where we have introduced the $2\times 2$
matrix

\begin{equation}
{\cal L}=\left(\matrix{0 & 1 \cr -1 & 0\cr}\right)\, .
\end{equation}

As we did in the Introduction we go to the Einstein metric
$g=e^{-\varphi}\jmath$ and obtain

\begin{eqnarray}
\label{slraction}
S={\textstyle\frac{1}{2}}\int d^6 x\sqrt{-g}\left[ -R(g)
  +{2\partial\lambda\partial\bar{\lambda}\over(\lambda-\bar{\lambda})^2}
  +{2\partial\kappa \partial\bar{\kappa }\over(\kappa -\bar{\kappa })^2}
\right.
\nonumber\\
\left.
-\kappa_2{\cal H}^T\hat{\cal M}{\cal H}
  +\kappa_1{\cal H}^T{\cal L}{}^{\ *}{\cal H}\right]\, .
\end{eqnarray}

\noindent The complex scalars $\lambda, \kappa$ are

\begin{eqnarray}
\kappa&=&\kappa_1 + i\kappa_2
={\textstyle\frac{1}{8}}D+{\textstyle\frac{3}{4}}ie^{2G}\, , \nonumber
\\ \lambda &=& \lambda_1 + i \lambda_2 = \ell + i e^{-\varphi}\, ,
\end{eqnarray}
with

\begin{equation}
\varphi= \bar G + \bar\varphi\, ,\hskip .5truecm
2G=\bar{G}-\bar\varphi\, .
\end{equation}

There are two $SL(2,\R)/U(1)$ scalar cosets in the action
(\ref{slraction}) and correspondingly there are two $SL(2,\R)$
symmetries of the equations of motion.  One of them is the original
$SL(2,\R)_{\rm IIB}$ symmetry of the NSD type~IIB action
(\ref{eq:stringynonselfdualaction}).  Note that $G$ (not $\bar{G}$) is
$SL(2,\R)_{\rm IIB}$ invariant.  The second is the e-m $SL(2,\R)_{\rm
EM}$ duality of the two-form potentials we were looking for, and it is
only a symmetry of the equations of motion.  It acts on $\kappa$ and
${\cal H}$ as follows:

\begin{eqnarray}
\kappa^{\prime} & = & {p\kappa +q\over r\kappa +s}\,,\nonumber\\
{\cal H}_{\mu\nu\rho}^{\prime} & = & (r\kappa_1 +s){\cal H}_{\mu\nu\rho}
+r\kappa_2 {\cal L}\hat{\cal M}{}^{\ *}{\cal H}_{\mu\nu\rho}\,,
\end{eqnarray}

\noindent with $ps-qr=1$.  Note that these $SL(2,\R)_{\rm EM}$
transformations are similar in form
to the $S$ duality of the heterotic string compactified to four
dimensions\footnote{See {\it e.g.}~the review of \cite{Se2}}, with
vector fields replaced by two-form fields and with the axion/dilaton
field replaced by $\kappa$\footnote{This is the reason that we use
the name $SL(2,\R)_{\rm EM}$.}.

We finally recall that the complete noncompact symmetry group of $D=6$
type~IIA supergravity \cite{Ta1} is $SO(5,5)$.  In this section we have
dimensionally reduced a truncated version of the type~IIB theory, using
the special ansatz (\ref{ansatzspecial}).  In this way we recovered a

\begin{equation}
SO(2,2) \equiv SL(2,\R)_{\rm EM} \times SL(2,\R)_{\rm IIB}
\end{equation}

\noindent noncompact symmetry of the equations of motion.
In the next section we will apply this symmetry to
construct a class of dyonic five-brane solutions.


\section{Dyonic Five--Branes}

Using the solution-generating transformations constructed in the
previous section, it is now straightforward to construct ten-dimensional
dyonic five-brane solutions $5_{(10)\rm d}$.  To this end we apply the
most general $SL(2,\R)_{\rm IIB} \times SL(2,\R)_{EM}$ transformation,
with parameters $a,b,c,d\ (ad-bc=1)$ and $p,q,r,s\ (ps-qr=1)$,
respectively, to the $5_{(10){\rm m}}$ solution given in
(\ref{5brane}).  The result is given by\footnote{All results in this
section are ten-dimensional and in string frame.  We omit all hats.}

\begin{eqnarray}
ds^2 &=& A\left[(dx^0)^2 - (dx^1)^2\right] -B(dx^m)^2
-Ae^{2C}(dx^a)^2\,, \nonumber\\
\left(
\begin{array}{c}
{\cal H}^{(1)}\\
       \\
{\cal H}^{(2)}
\end{array}
\right)
&=&
\left(
\begin{array}{c}
asH -{\textstyle\frac{3}{4}}bre^{-2C}{}^{\ *}H\\
                                   \\
csH -{\textstyle\frac{3}{4}}dre^{-2C}{}^{\ *}H
\end{array}
\right)
\,,\\
e^{-\varphi} &=& {e^{-C}\over a^2+b^2e^{-2C}}\,,\nonumber\\
\ell &=& {ac+bde^{-2C}\over
          a^2 + b^2e^{-2C}}\,,\nonumber\\
D_{mnpq} &=& {\textstyle\frac{1}{3}}\epsilon_{mnpq}{qs
+{\textstyle\frac{9}{16}}pre^{-2C} \over
s^2+{\textstyle\frac{9}{16}}r^2e^{-2C}}\,.\nonumber
\end{eqnarray}

\noindent where the functions $A,B$ and $H_{abc}$ are functions of $C$

\begin{eqnarray}
A &=& {\sqrt {a^2+b^2e^{-2C}}}{\sqrt
{s^2 +{\textstyle\frac{9}{16}}r^2e^{-2C}}}\,,\nonumber\\
B &=& {{\sqrt {a^2 + b^2e^{-2C}}}\over {
\sqrt {s^2 + {\textstyle\frac{9}{16}}r^2e^{-2C}}}}\,,\\
H_{abc} &=&
{\textstyle\frac{2}{3}}\epsilon_{abcd}\partial^dC\,,\nonumber\\
\end{eqnarray}

\noindent and $C$ depends only on the $x^{a}$'s and satisfies
$\Box e^{2C}=0$.

A characteristic feature of the above dyonic fivebrane solutions is that
non-zero RR fields are needed in order for the solution to carry
electric as well as magnetic charge.  Setting the RR axion and the other
RR fields $\ell$ and $D_{mnpq}$ equal to zero, leads to a purely
electric or purely magnetic solution.

The above family of dyonic fivebrane solutions contains the known four
purely electrically or magnetically charged fivebrane solutions (see
Introduction) as special cases.  First of all, for $a=d=p=s=1,
b=c=q=r=0$ (unit transformation) we recover $5_{(10){\rm m}}$ which we
call $5^{(1)}_{(10){\rm m}}$ here\footnote{The superscript $(1), (2)$
indicates that the charge of the solution is carried by the NS-NS or RR
axion, respectively.}.

Secondly, the $(\Z_2)_{\rm IIB}$ transformation $b=p=s=1, c=-1, a=d=q=r=0$,
when acting on the $5^{(1)}_{(10){\rm m}}$ solution,
leads to the electrically charged solution $5_{(10){\rm
e}}^{(2)}$\footnote{Note that according to our point of view this solution
is of electric
character because the string coupling constant $e^{\varphi}$ is small
($e^{-2\varphi}$ is singular). This property is due to the fact that the axion
involved is a RR axion which from the string theory point of view
is a (world-sheet) soliton. {\it All} solutions have the property that
the dilaton corresponding to the electrically (magnetically)
charged solutions satisfies $\Box e^{-2\varphi} = 0\ (
\Box e^{2\varphi}=0)$. In this way the electric and magnetic solutions
are always connected via a strong/weak coupling duality.}:

\begin{equation}
{\bf 5^{(2)}_{(10){\rm e}}}\left\{
\begin{array}{rcl}
ds^2 &=& e^\varphi\left[ (dx^0)^2 - (dx^1)^2 - (dx^m)^2\right]
- e^{-{\varphi}}(dx^a)^2\,,\\
{\cal H}^{(2)}_{abc} &=&
{\textstyle\frac{2}{3}}\epsilon_{abcd}\partial^d {\varphi}\,.\\
\varphi &=& \varphi(x^a)\, , \hspace{1cm} \Box e^{-2\varphi}=0\, .
\nonumber
\label{fivem2}
\end{array}\right.
\end{equation}

Thirdly, the $5_{(10)\rm e}^{(1)}$ solution in Eqs.~(\ref{FS}) is
obtained by applying the $(\Z_{2})_{\rm IIB}\times (\Z_{2})_{\rm EM}$
transformation $b=1,c=-1,q=-3/4,r=4/3,a=d=p=s=0$ on the $5_{(10){\rm
m}}^{(1)}$ solution.

Finally, by applying the $(\Z_2)_{\rm EM}$ transformation
$a=d=1,q=-3/4,r=4/3, b=c=p=s=0$ on the $5_{(10){\rm m}}^{(1)}$ solution
we obtain the second magnetically charged solution $5_{(10){\rm m}}^{(2)}$:

\begin{equation}
{\bf 5^{(2)}_{(10){\rm m}}}\left\{
\begin{array}{rcl}
ds^2 &=& e^{-\varphi}\left [ (dx^0)^2 - (dx^1)^2\right ] -
e^{\varphi}\left [(dx^m)^2
+ (dx^a)^2\right ]\,,\\
B_{01}^{(2)} &=& -e^{-2\varphi}\,,\nonumber\\
\varphi &=& \varphi(x^a)\, ,\hspace{1cm} \Box
e^{2\varphi}=0\, .\nonumber
\label{fivee2}
\end{array}\right.
\end{equation}

Observe that the original e-m $\Z_2$-transformation discussed in
the introduction that transforms the $5_{(10){\rm m}}^{(1)}$ solution
into the $5_{(10){\rm e}}^{(1)}$ solution, is the product of a
$(\Z_2)_{\rm IIB}$ and a $(\Z_2)_{\rm EM}$.

Finally, the $5_{(10){\rm m}}^{(1)}$ and $5_{(10){\rm e}}^{(1)}$
solutions are also solutions of the heterotic superstring whereas the
$5_{(10){\rm m}}^{(2)}$ and $5_{(10){\rm e}}^{(2)}$ solutions are
solutions of the type~I superstring \cite{Da2}.  These particular
solutions have been used in \cite{Da2} to confirm the ten-dimensional
duality between the heterotic and type~I superstring.

\section{Conclusions and Outlook}

In this letter we have constructed dyonic five-brane solutions in ten
dimensions by applying a solution-generating $SL(2,\R)_{\rm IIB}\times
SL(2,\R)_{\rm EM}$ $S$~duality transformation.  The ten-dimensional dyonic
five-brane solutions can also be interpreted as six-dimensional dyonic
string solutions.  It is interesting to compare our results with the
six-dimensional dyonic string solution that was recently constructed in
\cite{Du7}.  Our solution differs from that of \cite{Du7} in the
following two respects.  First of all, the dyonic string of \cite{Du7}
contains no RR fields whereas our solution does.  For instance, our
solution contains two axions while the one of \cite{Du7} contains one.
Secondly, the solution of \cite{Du7} is a heterotic solution that breaks
3/4 of the spacetime supersymmetries.  Our solution is a type~II
solution that breaks 1/2 of the type~II spacetime supersymmetries.  This
is necessarily so because the purely electrically or magnetically
charged five-brane has this property and we know that the $SL(2,\R)_{\rm
IIB}\times SL(2,\R)_{\rm EM}$--transformation, viewed as a noncompact
symmetry of six-dimensional supergravity is consistent with the full set
of type~II supersymmetries.

Although we give in this letter explicit results only for the five-brane
in ten dimensions, we expect that the techniques we use can be applied
to more general cases.  To explain the basic idea\footnote{The discussion
below is similar to that of \cite{Du8}.}, consider a
supergravity theory containing a metric $g_{\mu\nu}$, dilaton $\phi$ and
a $(p+1)$--form gauge field $B_{\mu_1\cdots \mu_{p+1}}$\footnote{
We only consider $p$-brane solutions where the charge is carried
by $B_{\mu_1\cdots \mu_{p+1}}$.  In particular, we do not consider here
$p$-brane solutions where the charge is carried by a vector component of
$g$ and/or $B$.}.  The Lagrangian for these fields takes a standard
form, as in \cite{Du1}. From the general analysis of \cite{Du1} it
follows that this theory has an elementary $p$-brane solution $p_{\rm
e}$ and a solitonic $(D-p-4)$-brane solution $(D-p-4)_{\rm m}$.  We next
observe that the dual of a $(p+1)$--form field is again a $(p+1)$--form
field in $2(p+2)$ spacetime dimensions.  In order to allow for a $\Z_2$
$S$~duality transformation we therefore reinterpret the $(D-p-4)_{\rm
m}$--solution in $D$ dimensions as a $p_{\rm m}$--solution in $2(p+2)$
dimensions via a dimensional reduction over $D-2p-4$ spacelike {\it
worldvolume} directions.  Similarly, a dimensional reduction of the
$p_{\rm e}$--solution in $D$ dimension over $D-2p-4$ spacelike {\it
transverse} directions leads to a $p_{\rm e}$--solution in $2(p+2)$
dimensions.  Given the standard form of the Lagrangean in $D$
dimensions, and assuming a simple ansatz for the $D$-dimensional fields
that includes the $D$-dimensional $p_{\rm e}$ and $(D-p-4)_{\rm m}$
solutions (as in (\ref{ansatzspecial})),
one can show that the field equations
corresponding to the dimensionally reduced theory in $2(p+2)$ dimensions
are invariant under a $\Z_2$ $S$~duality transformation that maps the
$p_{\rm e}$ and $p_{\rm m}$ solutions into each other.  In summary, the
special case of the $p_{\rm e}$ solution in $D$ dimensions where there
are $(D-2p-4)$ extra abelian isometries in the transverse directions can
be viewed, via a $\Z_2$ $S$~duality transformation in $2(p+2)$
dimensions, as the purely ``electrically'' charged partner $(D-p-4)_{\rm
e}$ of the ``magnetically'' charged $(D-p-4)_{\rm m}$ soliton solution
in $D$ dimensions.

It is instructive to consider a few examples of the above general
analysis.  Consider for instance ten-dimensional type~IIA supergravity.
The theory contains a $1,2$-- and $3$--form gauge fields and therefore
has the following solutions (see {\it e.g.}~\cite{Du8} or
the table in \cite{To1}):

\begin{equation}
\bigl( 0_{\rm e}, 6_{\rm m}\bigr)\, ,\hskip .5truecm
\bigl(1_{\rm e}, 5_{\rm m}\bigr)\, , \hskip .5truecm
\bigl(2_{\rm e}, 4_{\rm m}\bigr)\, .
\end{equation}

\noindent Applying the above analysis for $D=10$ and $p=0,1,2$,
respectively, we see that all the elementary solutions can be
reinterpreted as purely ``electrically'' charged parners of the
``magnetically'' charged soliton solutions by a $\Z_2$ $S$~duality
transformation in $4,6$ and $8$ dimensions, respectively:

\begin{equation}
0_{\rm e} \rightarrow 6_{\rm e}\, ,\hskip .5truecm
1_{\rm e} \rightarrow 5_{\rm e}\, , \hskip .5truecm
2_{\rm e} \rightarrow  4_{\rm e}\, .
\end{equation}

\noindent Note that only the string/five-brane solution can also be
considered as a solution of the heterotic superstring.

Next, we consider type~IIB supergravity.  It contains a complex
$2$--form and a self--dual $4$--form gauge field.  The complex $2$--form
gauge field leads to the following solutions:

\begin{equation}
\bigl(1_{\rm e} + 1_{\rm e}, 5_{\rm m} + 5_{\rm m}\bigr)\, .
\end{equation}

\noindent In addition, the self--dual $4$--form gauge field leads to the
self--dual threebrane solution $3_{\rm em}$ of \cite{Ho2,Du2}.  By reducing to
six dimensions we find that

\begin{equation}
1_{\rm e} + 1_{\rm e} \rightarrow 5_{\rm e} + 5_{\rm e}\, .
\end{equation}

\noindent The self--dual $3_{\rm em}$ solution is special in the sense
that our general formulae given above lead to a $\Z_2$--duality
transformation in ten dimensions itself.  However, since type~IIB
supergravity is already self--dual there is no such
$\Z_2$--transformation\footnote{See, however, section
\ref{sec-iibaction} where a related $\Z_2$--transformation, which is
present in the NSD theory, is mentioned.}.

Finally, we consider the case of eleven-dimensional supergravity.  There
is only one $3$--form in eleven dimensions which leads to an elementary
membrane $2_{\rm e}$ and a solitonic five-brane $5_{\rm m}$.  By
performing a $\Z_2$ $S$~duality transformation in $8$ dimensions we find
that

\begin{equation}
2_{\rm e} \rightarrow 5_{\rm e}\, .
\end{equation}

The results of this work suggest that in all examples given above,
the $\Z_2$ $S$~duality
transformation can be extended to an $SL(2,\R)$ transformation in a
relatively simple way.  The $SL(2,\R)$--transformations so obtained can
then be applied to construct dyonic $4,5$-- and $6$-brane solutions in
$D=10$ and a dyonic $5$-brane solution in $D=11$. It would be
interesting to construct these dyonic $p$--brane solutions and to
investigate their properties, like {\it e.g.} their singularity structure.

\bigskip

\noindent{\sl Note Added:}\ After this work was completed, an interesting
paper appeared \cite{Va1} where, for quite different purposes, also
the reduction of $D=10$ type~II supergravity on $T^4$ is considered.
The authors of \cite{Va1} reduce the type~IIA theory while in this letter
the reduction of the (truncated) type~IIB theory is considered.


\section*{Acknowledgements}

We thank Mees de Roo for useful discussions.
One of us (T.O.)~is extremely grateful to the hospitality, friendly
environment and financial support of the Institute for Theoretical
Physics of the University of Groningen.  The work of E.B.~has been made
possible by a fellowship of the Royal Netherlands Academy of Arts and
Sciences (KNAW).  The work of T.O. was supported by a European Union
{\it Human Capital and Mobility} program grant. The work of H.J.B.~was
performed as part of the research program of the ``Stichting voor
Fundamenteel Onderzoek der Materie'' (FOM).

\appendix


\end{document}